\documentclass[twocolumn]{aastex6}
\usepackage{amsmath}

\begin{document}

\title{Dust Attenuation Curves at z $\sim$ 0.8 from LEGA-C:\\ Precise Constraints on the Slope and 2175\AA\ Bump Strength}
\author{
Ivana Bari\v{s}i\'c$^1$,
Camilla Pacifici$^2$,
Arjen van der Wel$^{1,3}$,
Caroline Straatman$^3$,
Eric F. Bell$^4$,
Rachel Bezanson$^5$,
Gabriel Brammer$^6$,
Francesco D'Eugenio$^3$,
Marijn Franx$^7$,
Josha van Houdt$^1$,
Michael V. Maseda$^7$,
Adam Muzzin$^8$,
David Sobral$^9$,
Po-Feng Wu$^{10}$}

\thanks{$^1$Max-Planck Institut f\"ur Astronomie, K\"onigstuhl 17, D-69117, Heidelberg, Germany}
\thanks{$^2$Space Telescope Science Institute, 3700 San Martin Drive, Baltimore, MD 21218, USA}
\thanks{$^3$Sterrenkundig Observatorium, Department of Physics and Astronomy, Ghent University, Belgium}
\thanks{$^4$Department of Astronomy, University of Michigan, 1085 S. University Ave, Ann Arbor, MI 48109, USA}
\thanks{$^5$University of Pittsburgh, Department of Physics and Astronomy, 100 Allen Hall, 3941 OHara St, Pittsburgh PA 15260, USA}
\thanks{$^6$The Cosmic Dawn Center, Rockefeller Komplekset, Juliane Ma, 2100 K{\o}benhavn {\o}}
\thanks{$^7$Leiden Observatory, Leiden University, P.O.Box 9513, NL-2300 AA Leiden, The Netherlands}
\thanks{$^8$Department of Physics and Astronomy, York University, 4700 Keele St., Toronto, Ontario, Canada, MJ3 1P3}
\thanks{$^9$Department of Physics, Lancaster University, Lancaster LA1 4YB, UK}
\thanks{$^{10}$National Astronomical Observatory of Japan, Osawa 2-21-1, Mitaka, Tokyo 181-8588, Japan}


\email{barisic@mpia.de}

\begin{abstract}
We present a novel approach to measure the attenuation curves of 485 individual star-forming galaxies with M$_*$ $>$ 10$^{10}$ M$_{\odot}$ 
based on deep optical spectra from the VLT/VIMOS LEGA-C survey and multi-band photometry in the COSMOS field. Most importantly, we find that the attenuation 
curves in the rest-frame $3000-4500$\AA\ range are typically almost twice as steep as the Milky Way, LMC, SMC, and Calzetti attenuation curves, 
which is in agreement with recent studies of the integrated light of present-day galaxies. The attenuation at $4500$\AA\ and the slope strongly 
correlate with the galaxy inclination: face-on galaxies show less attenuation and steeper curves compared to edge-on galaxies, suggesting that 
geometric effects dominate observed variations in attenuation. Our new method produces $2175$\AA\ UV bump detections for 260 individual 
galaxies. Even though obvious correlations between UV bump strength and global galaxy properties are absent, strong UV bumps are most often seen in face-on, lower-mass galaxies (10 $<$ log$_{10}$(M$_*$/M$_{\odot}$) $<$ 10.5) with low overall attenuation. Finally, we produce a typical 
attenuation curve for star-forming galaxies at $z\sim0.8$; this prescription represents the effect of dust on the integrated spectral energy 
distributions of high-redshift galaxies more accurately than commonly used attenuation laws.
\end{abstract}

\section{Introduction}
Examining dust properties of galaxies improves our understanding of the evolution of galaxies through cosmic time. Yet, properly addressing and interpreting properties of dust presents a very challenging task. One of the techniques applied to tackle this challenge is to investigate the effect of dust on stellar light at different wavelengths, which was done first through examination of extinction curves \citep[e.g.][]{savage75, fitzpatrick86, fitzpatrick88, fitzpatrick90}. Extinction curves describe line of sight effects of the influence of dust on stellar light, but in order to understand the global effect of dust in a galaxy it is necessary to measure attenuation (absorption and scattering) of the integrated light. The first step toward understanding attenuation is to determine the wavelength dependence of the attenuation curve. Thus far, dust attenuation studies faced limitations due to the inability to directly measure the intrinsic stellar spectrum. This precludes the determination of the attenuation curve without strong degeneracies with stellar population properties such as age and metallcity.


Before discussing how we address this issue, let us first summarize how the description of extinction and attenuation curves have evolved over time: from e.g. \cite{savage75} who applied a linear ($\lambda^{-1}$) extinction curve baseline term, to those involving higher order polynomial applied by studies that followed \citep[e.g.][]{fitzpatrick86, cardelli89}. Independently of the extinction curve baseline choice, Milky Way studies included an additional term (i.e. Drude profile) to describe the profile and strength of the prominent UV bump feature. As more observations of other local and low-redshift galaxies emerged, attenuation curve studies focused primarily on the baseline, as the contribution of the UV bump was not initially observed \citep[e.g.][]{calzetti00} 
Later on, dust attenuation curves of local and high-redshift galaxies have often been described by a high-order polynomial \citep[e.g.][]{buat12, battisti17} or a modified power-law version of Calzetti dust law \citep[e.g.][]{noll09}, in combination with a Drude profile to account for a possible presence of the UV bump feature. 
These low- and high- redshift attenuation curve studies relied on the locally derived extinction curves for Small Magellanic \citep[e.g.][]{prevot84, gordon98} and Large Magellanic Clouds \citep[e.g.][]{clayton85} in order to interpret and compare measured features \cite[e.g.][]{gordon03, munoz04}.
Understanding the origin of the UV bump \citep{stecher65, savage75} and its properties is important to interpret the physical properties of those galaxies that demonstrate this feature, and to explore the evolution of these properties through cosmic time. A number of theoretical and laboratory based studies have been conducted over the last few decades in attempt to explain the origin of the UV bump feature, and more recent findings suggest PAH molecules as a promissing carrier candidate of the 2175\AA\ feature \citep{joblin92, beegle97, steglich10}.
The UV bump, together with the attenuation curve slope, which describes the reddening of the attenuation curve, have been used to characterize properties of the attenuation by dust \citep[e.g.][]{burgarella05, buat12, kriek13, battisti17, tress18, narayanan18, salim18}. 
However, studies so far suggest that the attenuation curve measurement is influenced by the geometrical effects, making the derivation of the dust properties more difficult \citep{wittgordon, pierini04, tuffs04, panuzzo06, chevallard13, seondraine}.

Up until now, dust attenuation studies in both the local and high redshift universe have mostly relied on deriving spectral energy distribution (SED) fits based on the observed multi-band photometry to recover the information about the attenuation in galaxies \citep[e.g.][]{buat12, gordon16, battisti17, salim18}, and some made use of Balmer decrement corrections from emission line spectroscopy \citep[e.g.][]{reddy15, battisti16, shivaei20}. For example, \cite{reddy15} found a steep attenuation law for a sample of z $\sim$ 2 galaxies compared to \cite{calzetti00}. In addition, \cite{kriek13} sampled the SED by combining narrow- and medium- band photometry of galaxies at similar redshifts, producing stacked high-resolution pseudo-spectra, providing significant evidence for the existence of the UV bump at z $\sim$ 2. 
\cite{scoville15} find evidence for the presence of the UV bump in their sample of high redshift z = 2 -- 6 galaxies. 
Finally, constraints on attenuation can also be inferred from the relation between the UV slope and the infrared excess, which has been done at various redshifts \citep[e.g.][]{panuzzo06, reddy06, capak15, bouwens16, salmon16, barisic17, bourne17, cullen17, faisst17, lofaro17, fudamoto17, mclure18, reddy18, wang18, alvarez19}.

The key point of the preceding overview is that previous studies all rely on photometry to jointly model the attenuation-free stellar continuum and the attenuation. However, stellar continuum spectroscopy can constrain the attenuation-free stellar spectrum with much better precision than photometry. Moreover, since the information of the stellar population is encoded in absorption lines that span a very small range in wavelength, attenuation does not greatly affect the interpretation of an absorption line spectrum. If one can reconstruct the attenuation-free stellar spectrum with the aid of such continuum spectroscopy, then the comparison with photometry produces a direct measurement of the attenuation curve. This is the method we develop in this paper, which is inspired and enabled by the Large Early Galaxy Astrophysics Census \citep[LEGA-C,][]{vdw16} survey that, for the first time, produces sufficiently deep continuum spectra of galaxies out to z $\sim 1$. 
The main goal of this work is to provide a more accurate description of the attenuation curve for the integrated light of galaxies.
This will help efforts to interpret the attenuation of distant galaxies in terms of both schematic dust distribution models and full radiative transfer-based mock observations (see \cite{salim20} for an excellent review of the current state of the art).
We examine the diversity in dust attenuation properties among our galaxies, with the main focus on the strength of the UV bump feature and the slope of the attenuation curve. Furthermore, we explore the dependence of these properties on galaxy orientation, specific star-formation rate (sSFR) and stellar mass M$_*$.
 
 The outline of this paper is as follows: in the following section 2 we introduce the data set used in this study, and the choice of attenuation curve parametrization. In section 3 we present a new attenuation curve prescription and discuss the results. A summary of this work is given in section 4. 

\section{Data}

\subsection{The LEGA-C Survey}

The Large Early Galaxy Astrophysics Census \citep[LEGA-C;][]{vdw16}, an ESO public spectroscopic survey, conducted between 2014 and 2018 with the VIMOS spectrograph at the Very Large Telescope, was devised to obtain high signal-to-noise (S/N $\sim$ 20\AA$^{-1}$) and high resolution \citep[R $=$ 3500;][]{straatman18} optical continuum spectra of high redshift galaxies \citep{vdw16}. The survey gathered continuum spectra of $>$ 3000 K-band selected galaxies from the UltraVISTA survey \citep{muzzin13} within the 1.62 square degree region of the COSMOS field at a redshift range between 0.6 $<$ z $<$ 1, covering the wavelength range between 6300\AA\ $\lesssim$ $\lambda$ $\lesssim$ 8800\AA\ .
This study utilizes the most recent Data Release II sample \citep{straatman18}, comprised of a total of 1989 galaxies.

We make use of spectroscopic redshifts and axis ratio. 
Additionally, we also make use of the observed multi-band photometry from the UltraVISTA photometric catalog \citep{muzzin13}. SFR values are derived from UV and 24$\mu$m photometry, following \cite{whitaker12} relation, while stellar mass M$_*$ estimates are obtained following \cite{pacifici12} (Pacifici et al. 2020, in prep).
For further details on data reduction and the most recent data release we refer to \cite{straatman18}. 

\begin{figure}
    \centering
    \includegraphics[width=\linewidth]{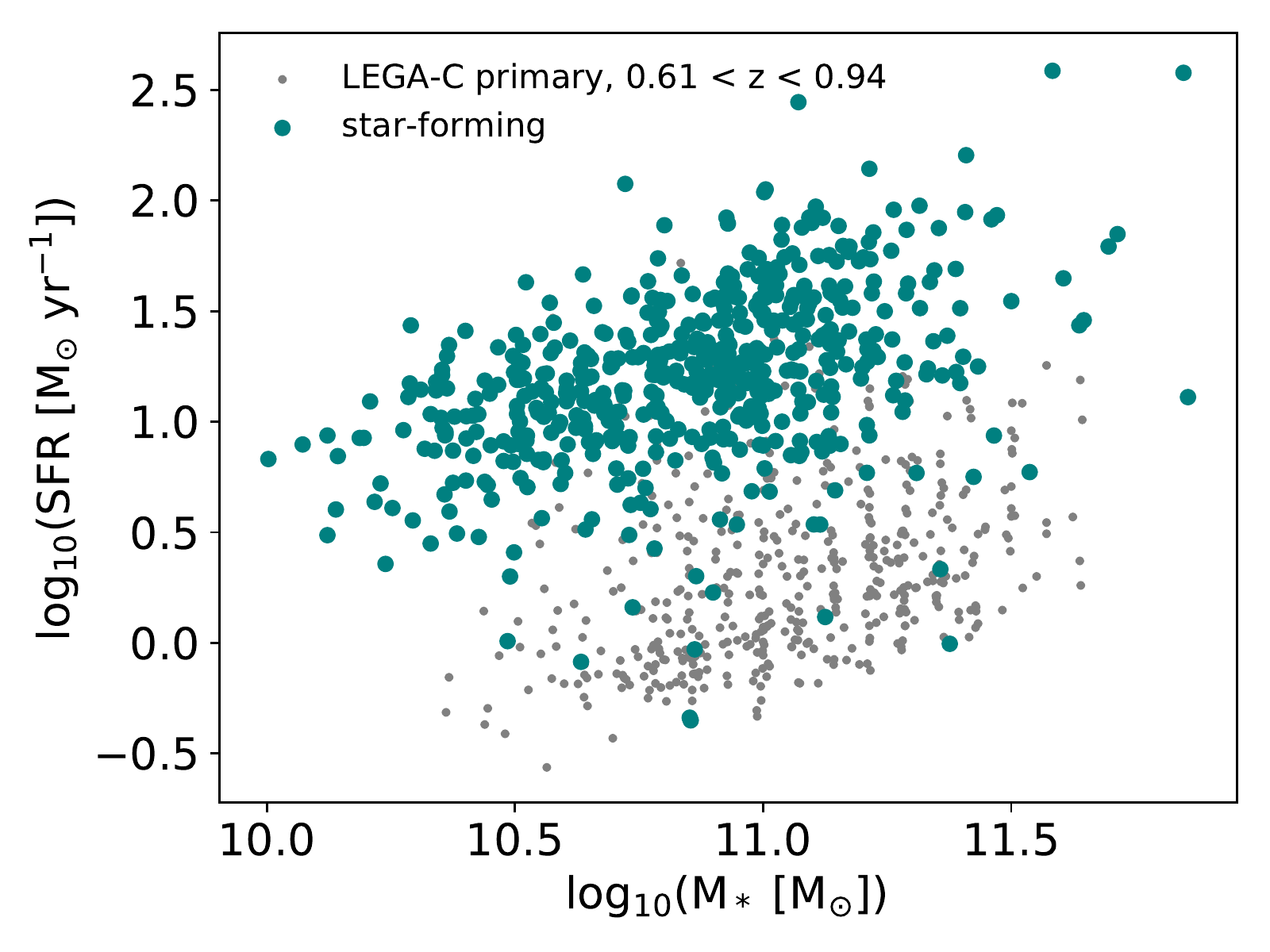}
    \caption{Star-formation rate vs stellar mass M$_*$ of 0.61 $<$ z $<$ 0.94 LEGA-C Data Release II sample shown with grey symbols. The selected sample of star-forming galaxies is shown with teal colored circles.}
    \label{fig:sample}
\end{figure}

We select 
UVJ color\footnote{Rest-frame colors are calculated using EAZY \citep{brammer08} based on UltraVISTA photometry \citep{muzzin13}.} based star-forming galaxies in the redshift range 0.61 $<$ z $<$ 0.94, with `use' flag $=$ 1 \citep[see][]{straatman18} and available intrinsic stellar spectra measurement. We target this redshift range to achieve coverage in the UV bump region with at least one ($u-$ band) measurement. As the typical width of UV bump is 350\AA\ \citep{noll09}, in the chosen redshift range the $u-$ band measurement probes the UV bump over a wavelength range of 2175\AA\ $\pm$ 200\AA. These criteria secure good observed continuum stellar spectra quality and enable us to consistently model the strength of the UV bump feature with available photometric coverage. The combination of this criteria yields a sample of 524 star-forming galaxies shown with pink and teal colored symbols in Figure \ref{fig:sample}. A small fraction of star-forming galaxies in our selected sample has low SFR based on their infrared luminosity, however, this does not affect our results.

\begin{figure*}
    \centering
    \includegraphics[width=0.7\linewidth]{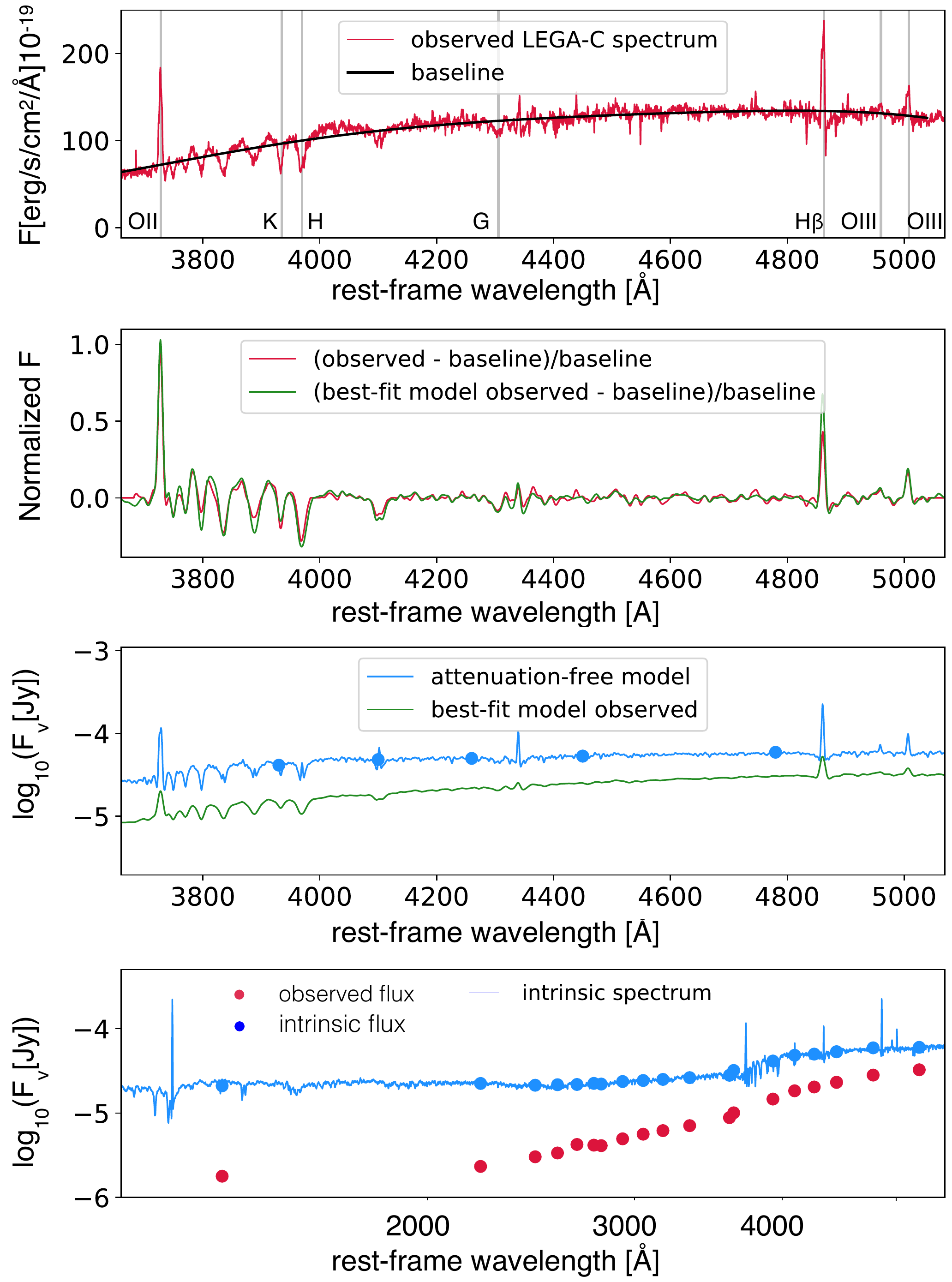}
    \caption{ Illustration of our fitting methods. {\bf{1st panel:}} Example of an observed LEGA-C spectrum (red curve, galaxy ID: 131198, {\textbf{z = 0.73}}) together with a baseline spectrum (black curve); {\bf{2nd panel:}} The normalized LEGA-C spectrum (red) and the corresponding best-fitting model spectrum (green). {\bf{3rd panel:}} Best-fit model spectrum including attenuation (green), and attenuation-free (blue) in original units (not normalized); {\bf{4th panel:}} The attenuation-free intrinsic spectrum over a broader wavelength range (blue curve) with synthensized photometric data-points (blue circles), and the observed photometric data points (red circles). The difference is our attenuation estimate.}
    \label{fig:steps}
\end{figure*}

\begin{figure*}
    \centering
    \includegraphics[width=\linewidth]{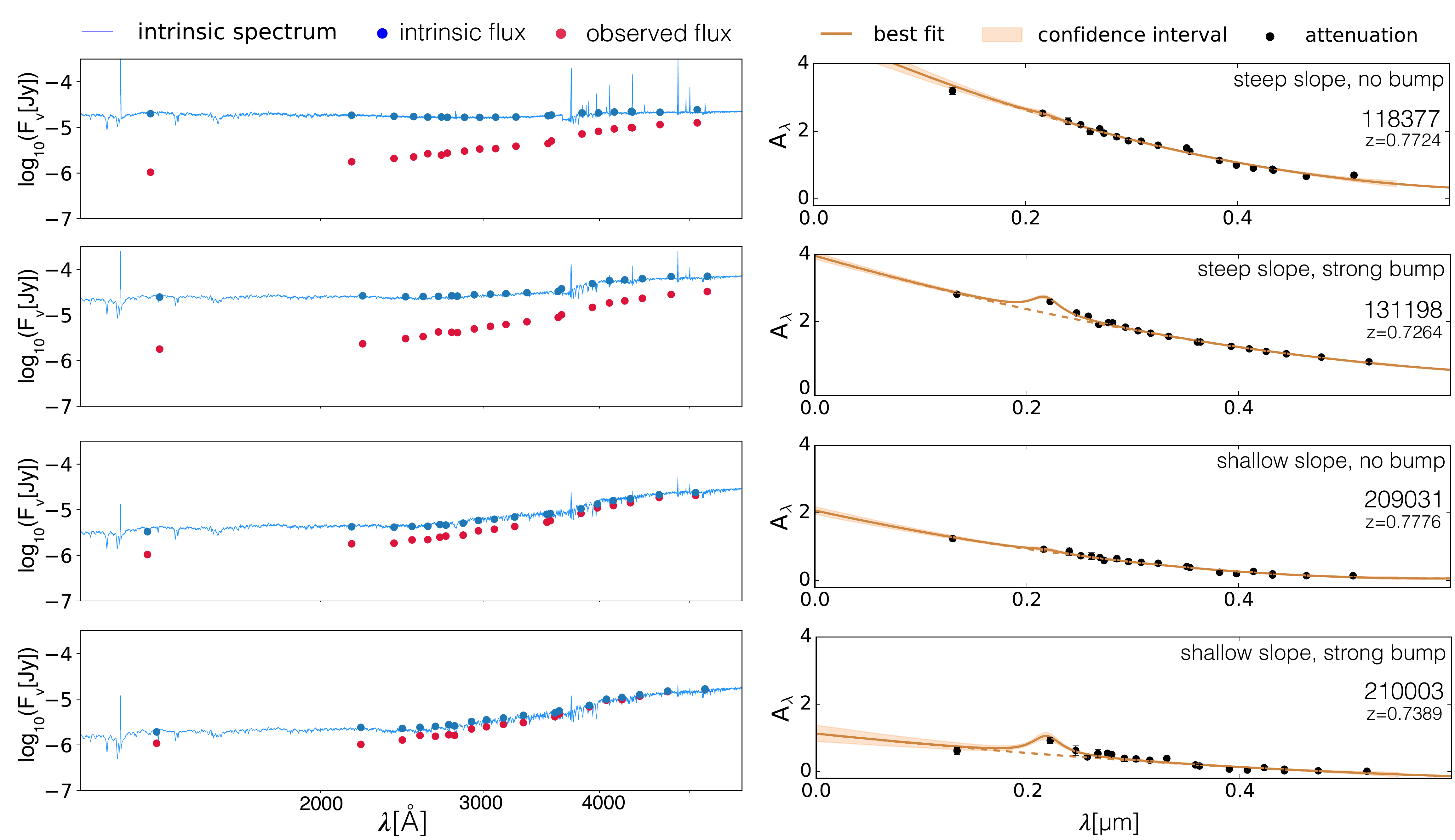}
    \caption{\textbf{Left:} Observed (red circles) and attenuation-free model (blue circles) flux density values, 
    together with attenuation-free model spectra (blue curve). \textbf{Right:} Attenuation data points together with best fit attenuation curves using a combination of the second order polynomial and the Lorentzian function to describe the bump feature, with 68\% confidence range. Error bars are smaller than the symbols. We show examples of steep/shallow attenuation curve with no bump detected (118377 and 209031) and prominent bump feature (131198 and 210003) respectively.}
    \label{fig:AC}
\end{figure*}

\subsection{Construction of Attenuation-Free Stellar Model Spectra}
\label{spectra}

The advantage of the deep and high-resolution LEGA-C spectra lies in the ability to estimate the stellar spectrum in a manner that is far less sensitive to attenuation than the usual approach to infer the stellar spectrum from photometry. The latter unavoidably leads to a large degree of degeneracy between the redness of the observed SED due to stellar population properties (age, star-formation history, metallicity) and reddening. 

Our approach is an adapted version of the methodology developed by \cite{pacifici12, pacifici16}; it is illustrated for one example galaxy in Figure \ref{fig:steps}. The first step is to construct a baseline spectrum from the observed high-resolution spectrum by taking its running median with a box size of 180$\AA$ (top panel of Figure \ref{fig:steps}). This smooth component accounts for the global shape of the spectrum that includes not only the effects of average reddening, but also imperfections in the flux calibration. We produce a normalized spectrum by subracting the baseline and then dividing by it: $N = (O - B) / B$, where $O$ is the observed spectrum and $B$ is the baseline (see second panel of Figure \ref{fig:steps}). This normalized spectrum contains the crucial information from which the star-formation history is inferred. This information consists of the combined effect of dozens of ion species in stellar atmospheres -- a few of which are labeled in the top panel of Figure \ref{fig:steps} -- that blend together and produce the integrated absorption line spectrum.

The synthetic model spectra constructed by \cite{pacifici16} are normalized in the same manner, with a different baseline spectrum that is calculated for each individual model spectrum. The library of model spectra are based on \cite{bruzual03} single-stellar population models combined with physically motivated star formation histories drawn from cosmological simulations are combined with the \cite{bruzual03}. It includes a prescription for nebular emission lines \cite{charlot01} and treats attenuation with a version of the two-component \cite{charlot00} dust model following variations in \cite{chevallard13} (for details see \cite{pacifici12}). It is important to note that the effect of global attenuation is removed from the model spectra through the normalization procedure: only the 2nd-order effect of differential attenuation of young and old stars affects the normalized model spectra. 
Also note that the emission lines only cover a few percent of the full wavelength range of the spectrum; as such they have little effect on the underlying stellar spectrum.

Both model and observed spectra are convolved with a Gaussian kernel in order to match a fixed resolution of $\sigma$ = 250~km~s$^{-1}$. The entire library of normalized model spectra is then compared to the observed normalized spectrum and the best-fitting model is chosen to represent the galaxies' SED\footnote{This procedure precludes a rigorous analysis of the uncertainty; we devise an empirical estimate of the uncertainty on the resulting attenuation curves in Sec 3.2.}. The third panel of Figure \ref{fig:steps} shows the original, non-normalized best-fitting model spectrum (with attenuation as prescribed by the \cite{charlot00} dust model) as well as the attenuation-free (intrinsic) model spectrum. The latter is used to synthesize photometric data points that are compared with observed photometry over a much broader wavelength range (bottom panel of Figure \ref{fig:steps}) than covered by the LEGA-C spectra.  The difference between the two produces our attenuation curve. Note that while the \cite{charlot00} dust model is used when fitting the spectra, it no longer plays a role in inferring our attenuation curve.

Since the attenuation slope measurement sensitively depends on where the attenuation reaches 0, we normalize the intrinsic, dust-free model spectra so that attenuation is forced to be zero at 3.3$\mu$m by linearly extrapolating from the UltraVISTA $J-$ and $Ks-$ band photometry in inverse wavelength space 1/$\lambda$. In principle, IRAC photometry should be more suitable for this normalization, as it is closer in wavelength, but the difficulty in matching photometry due to the large IRAC PSF precludes us from doing so with sufficient accuracy.
Several examples of the intrinsic stellar spectra, together with the intrinsic and observed flux density values at corresponding UltraVISTA bands can be seen in the left panels of Figure \ref{fig:AC}. In each case the difference between the attenuation-free, renormalized model and the observed broad-band photometry provides us with the attenuation estimate (as shown in right panels of Figure \ref{fig:AC}). 

\subsection{Attenuation Curve Fitting}
\label{attenuation}
The UltraVISTA catalog contains flux density values in 30 photometric bands, and our attenuation curve determination is based on a subset of those.
GALEX $FUV-$ band is excluded as it covers rest-frame Lyman-$\alpha$ forest in the selected redshift range (0.61 $<$ z $<$ 0.94). We choose to fit the attenuation curve in 20 photometric bands including: GALEX $NUV-$, CFHT Megaprime $u-$, SUBARU Suprime-Cam intermediate and broad ($B-$, $V-$, $g^+-$, $r^+-$, $i^+-$, $z^+-$) band filters -- included in Figure \ref{fig:AC}. The observed attenuation is obtained via:

\begin{equation}
    \nonumber A = -2.5 \cdot \log_{10} \bigg(\frac{f_{obs}}{f_{int}}\bigg)
\end{equation}

\noindent where f$_{obs}$ and f$_{int}$ are the observed and intrinsic flux density values. Attenuation in the GALEX $NUV-$ band puts a constraint on the blue part of the attenuation curve as it is the only data point at shorter wavelength than the UV bump (covered by the $u-$ band). 

Given the diversity in the literature when it comes to the choice of the attenuation curve parametrization 
to be fitted to the observed attenuation \citep[e.g.][]{noll09, battisti17}, we opt for the second order polynomial to describe the baseline of the attenuation curve, in a linear combination with the Lorentzian function\footnote{http://mathworld.wolfram.com/LorentzianFunction.html} to describe the UV bump feature: 

\begin{equation}
 A(\lambda) = a\lambda^2 + b\lambda + c + \frac{\alpha}{\pi}\frac{0.5 \Gamma}{(\lambda - \lambda_0)^2 + (0.5 \Gamma)^2} 
 \label{eq:param}
\end{equation}

\noindent Here $\lambda$ is wavelength in $\mu$m, $\Gamma$ is the width of the bump feature \citep[0.035$\mu$m,][]{noll09}, $\lambda_0$ is the central wavelength of the bump (0.2175 $\mu$m), and $B$ is the amplitude parameter to scale the Lorentzian function.  
The parameters $a$, $b$, $c$ are free fitting parameters of the second order polynomial, as is the amplitude $\alpha$ of the Lorentzian function. We define the strength of the UV bump feature in the following way:

\begin{equation}
\label{bump}
 B = \frac{A(2175\AA) - A_{base}(2175\AA)}{A_{base}(2175\AA)}    
\end{equation}

\noindent where A$_{base}$($\lambda$) is the baseline value of the second order polynomial. 

Largest variations among the attenuation curves are best seen in the UV range. Still, a number of attenuation curve studies over the years choose to define the slope of the attenuation curve in the optical wavelength range. The drawback of this choice is that it is not best in representing the slope of the whole attenuation curve. However, to be consistent with majority of the conducted studies, we choose to define the slope of the attenuation curve in the optical wavelength range as well -- opting for a non traditional R(4500\AA):

\begin{equation}
\label{slope}
     R(4500\AA) = \frac{A(4500\AA)}{A(3000\AA) - A(4500\AA)}
\end{equation}

\noindent We choose this attenuation curve slope definition over the typical R$_V$\footnote{ R$_V$ $\equiv$  A$_V$ / E(B-V) $\equiv$ A$_V$ / (A$_B$ - A$_V$) } since the second order polynomial in our parametrization A($\lambda$) often shows an inflection at $\approx$ 5500\AA, leading to non-informative R$_V$ values. The physical origin of this inflection is discussed below, when we make a comparison with the results from standard attenuation law prescriptions (e.g. modified Calzetti law).


Since our parametrization differs from the standard prescriptions, it is useful to calculate UV bump strength B and slope of the attenuation curve R(4500\AA) for the widely used attenuation/extinction laws. Applying our parametrization to the \cite{cardelli89} prescription for the Milky Way (MW) extinction curve we find B $\sim$ 0.48, and a slope R(4500\AA) $\sim$ 2.2, whereas for the \cite{calzetti00} attenuation law we find R(4500\AA) $\sim$ 2.4. Extrapolating from the data given in Table A4 in \cite{gordon03} we find R(4500\AA) $\sim$ 2.1 and 2.5 for the Small Magellanic Cloud (SMC) and Large Magellanic Cloud (LMC), respectively. 

\begin{figure*}
    \centering
    \includegraphics[width=\linewidth]{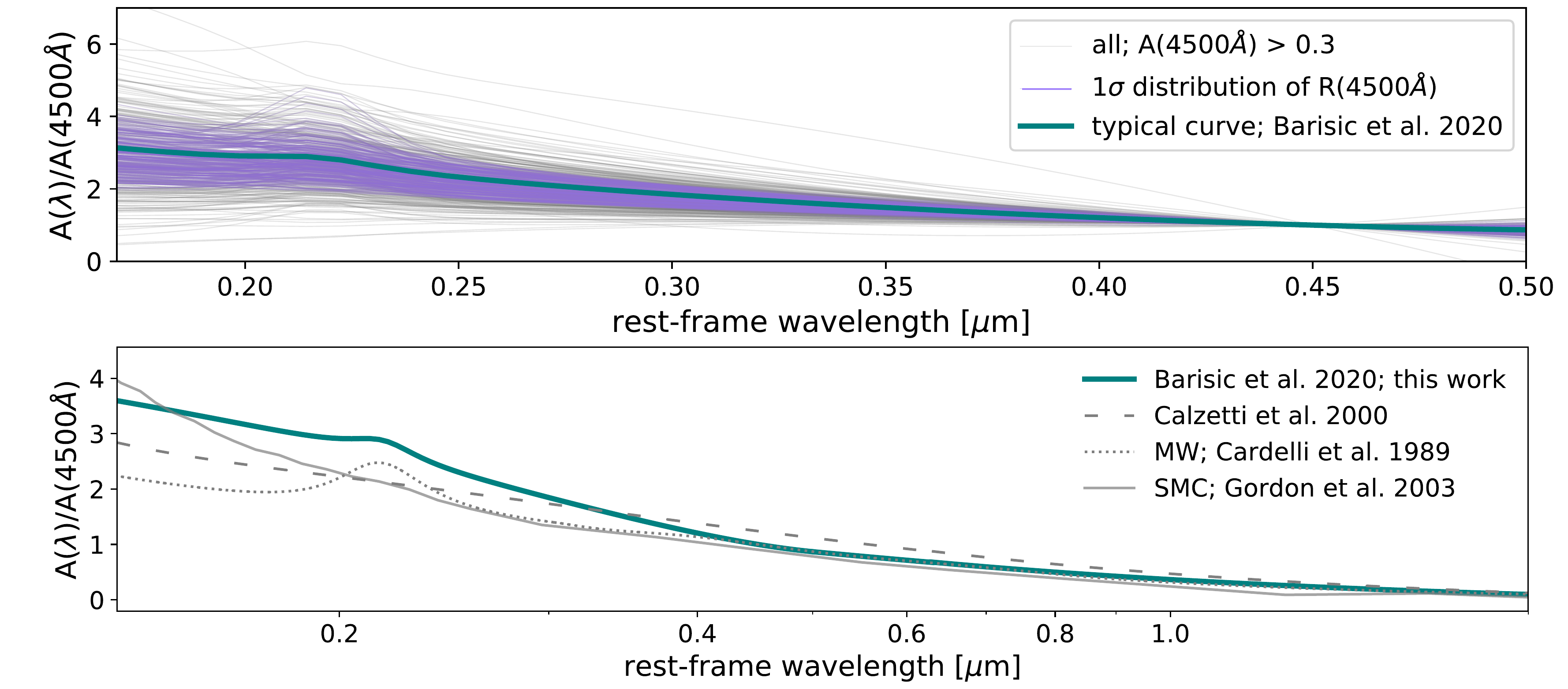}
    \caption{{\bf{Top:}} All of the attenuation curves (gray) from a subset of the selected sample (A(4500\AA) $>$ 0.3); representing the scatter -- we highlight attenuation curves (purple) of those galaxies within 1$\sigma$ of the slope R(4500\AA) distribution. The typical attenuation curve (teal) is also shown. 
    {\bf{Bottom:}} A typical attenuation curve for z $\sim$ 0.8 star-forming galaxies (solid curve) out to 0.5$\mu$m. At wavelengths longer than 0.5$\mu$m we show re-scaled \cite{calzetti00} attenuation curve (0.5$\mu$m -- 2.2$\mu$m). For comparison we also show Milky Way \cite{cardelli89} extinction curve re-normalized to 4500\AA. We also show the standard \cite{calzetti00} attenuation curve and Small Magellanic Cloud \citep{gordon03} extinction curve re-scaled by the same factor as Milky Way.}
    \label{fig:typical}
\end{figure*}


We fit our A($\lambda$) parametrization (see Equation \ref{eq:param}) to the observed attenuation in several steps. If any observed attenuation data point is $>$3$\sigma$ from the best-fit parametrization curve, it is excluded and the fit is performed again. As a next step, we make sure an optimal fit to the observed attenuation around the UV bump region is achieved. This is done by integrating the $u-$ band (together with $IB427-$ band at z $>$ 0.8) transmission curve \citep{muzzin13} with the most-recent best-fit to obtain the predicted attenuation data point. The difference between the predicted and the observed attenuation data point is then added to the observed data point, and the fit is performed again. We find two iterations leads to convergence. We prefer this approach of fitting data points to the attenuation curve over the much computationally slower approach of convolving the attenuation curve with all filter transmission curves to find the best fit. A variety of examples of the attenuation data points and the resulting attenuation curves is shown in the right panels of Figure \ref{fig:AC}.
Applying this fitting procedure, we reject\footnote{7\% of the total sample of 524 galaxies} galaxies with highly negative UV bump strength measurement (B $<$ - 0.2), and those with large uncertainties on the bump measurement (B$_{err}$ $>$ 0.2) 
as an indicator of the good fit, which gives us a sample of 485 galaxies (teal colored symbols in Figure \ref{fig:sample}). Out of 485 galaxies, 260 have the UV bump measured with 99\% confidence. 

Apart from the second order polynomial, we also explored several other parametrizations, including the most often applied -- the modified Calzetti dust law combined with the Drude profile, as given in e.g. \cite{salim18}. However, the issue we encountered while applying this particular parametrization is that the power-law fit provided poor fits for the majority of galaxies in our sample\footnote{Reduced $\chi^2$ $>$ 1.5 for $>$60\% of star-forming galaxies when using power-law, as opposed to $\sim$30\% when using a second order polynomial. {Reduced $\chi^2$ is by definition $\chi^2$ per degree of freedom, where degrees of freedom correspond to number of data points minus the number of free fitting parameters of the function}}. For this reason we adopted the second order polynomial as our default parametrization.
The main reason for this is that the galaxies in our sample generally have a broad age range in their stellar populations, and the dust attenuation is likely strongly age dependent. 
Old stellar populations with little dust attenuation dominate at longer wavelengths, while young stellar populations with more dust attenuation can dominate at shorter wavelengths. This leads to curvature in the attenuation that cannot be captured with a power law prescription, as it often ends up being too steep in the UV. For this reason studies often make use of two-component dust models {\citep[e.g.][]{charlot00, popescu00, tuffs04}.}
This phenomenon also strongly affects the inferred slope, which will be discussed further in Section \ref{res}.

\section{Results}
\label{res}

\subsection{Attenuation Curve Prescription}
As the existing parametrization prescriptions in the literature do not provide good quality fits for the observed attenuation of star-forming galaxies in our sample, we provide our own prescription. 
The following prescription is meant to reproduce the median slope R(4500\AA) and UV bump values of star-forming galaxies in the selected sample (R$_{median}$(4500\AA) $\sim$ 1.2 $\pm$ 0.31, B$_{median}$ $\sim$ 0.12 $\pm$ 0.07). To achieve this we select a subset of attenuation curves in a narrow slope range 1 $<$ R$_{median}$(4500\AA) $<$ 1.4, and simultaneously fit all of the attenuation data points, achieving a mean attenuation curve.
This prescription, represents a typical curve for star-forming galaxies at z $\sim$ 0.8, normalized at 4500\AA, and is defined in the wavelength range 0.13$\mu$m $<$ $\lambda$ $<$ 0.5$\mu$m: 

\begin{equation}
    \label{prescription}
    \frac{A(\lambda)}{A(0.45\mu m)} = 
    \begin{cases}
       ((14.780\pm0.303)\lambda^2 - (16.654\pm0.231)\lambda &  \\ + (5.498\pm0.044) + D_{\lambda}),   &  \\ 
      D_{\lambda} = \frac{0.017\pm0.001}{\pi}\frac{0.5\Gamma}{(\lambda - \lambda_0)^2 + (0.5\Gamma)^2}, & \\
      0.13\mu m < \lambda < 0.5\mu m   & 
    \end{cases}
\end{equation}

\noindent Reproduced slope and UV bump values following this prescription are 1.18 and 0.12 respectively. At wavelengths longer than 0.5$\mu$m we recommend the standard \cite{calzetti00} prescription, scaled by factor $N$ to match our prescription:

\begin{equation}
    A(\lambda) = 
    \begin{cases}
        \nonumber N \cdot (2.659(-2.156 + \frac{1.509}{\lambda} - \frac{0.198}{\lambda^2} + \frac{0.011}{\lambda^3}) + R_V),  & \\
        N = 0.194; 0.5\mu m < \lambda < 0.63\mu m    & \\  & \\

        N \cdot (2.659(-1.857 + \frac{1.040}{\lambda}) + R_V),    & \\
        N = 0.194; 0.63\mu m < \lambda < 2.2\mu m   & 
   \end{cases}
\end{equation}

Wavelength $\lambda$ used in both equations is in $\mu$m. In order to account for the variation in the slope of the attenuation curve, we offer a modified version of Equation \ref{prescription}:

\begin{align}
\label{modified}
    \begin{split}
      A'(\lambda) ={}& \frac{A(\lambda)}{A(0.45\mu m)} \cdot \Big( \frac{\lambda}{0.45} \Big) ^\delta 
    \end{split}\\
    \nonumber \\
    \begin{split}
      \nonumber  \delta ={}& \frac{log_{10}(R(4500\AA)/1.18)}{log_{10}(0.45/0.3)}
    \end{split} 
 \nonumber 
\end{align}


Upper panel in Figure \ref{fig:typical} presents all of the attenuation curves in the selected narrow slope range (gray) together with the typical curve following from the fit (teal) and its confidence interval. Teal solid curve in Figure \ref{fig:typical} shows our parametrization prescription up to 5000\AA, combined at longer wavelengths with re-scaled \cite{calzetti00} curve. For comparison we also show MW extinction curve \citep{cardelli89} which we have normalized to 4500\AA\ (dotted grey curve). The standard \cite{calzetti00} attenuation curve (dashed grey curve), and Small Magellanic Cloud as given by \citep{gordon03} (solid grey curve) are also shown, both of which have been re-scaled by the same factor as MW curve in order to preserve their mutual R$_V$ ratios.

Several recent studies have demonstrated a flatter attenuation curve slope in the optical to near-IR wavelength range as compared to a steep \cite{calzetti00} curve \citep[e.g.][]{chevallard13, lofaro17, trayford17, buat18, roebuck19}. However, most of the degeneracies between attenuation curves are at wavelengths shorter than 5000\AA. Still, a large number of attenuation curve studies continue using a modified \cite{calzetti00} law prescription. This justifies our Calzetti prescription recommendation at wavelengths longer than 0.5$\mu$m.
We note, however, that the Calzetti law prescription is only a suggestion and future studies are encouraged to apply the prescription of their choice at wavelengths longer than 5000\AA.
In the table below we provide normalized attenuation A($\lambda$)/A(4500\AA) values for our prescription:

\begin{deluxetable}{cccc}
\tabletypesize{\scriptsize}
\tablecaption{Values for the attenuation curve prescription (A($\lambda$)/A(4500\AA)) shown with teal solid curve in Figure \ref{fig:typical}}
\tablewidth{0pt}
\tablehead{
\colhead{$\lambda$[$\mu$ m]}     & \colhead{A($\lambda$)/A(4500\AA)}   &   \colhead{$\lambda$[$\mu$ m]}     & \colhead{A($\lambda$)/A(4500\AA)} }
\startdata
0.13    &   3.595 $\pm$ 0.01   &   0.5     &   0.867 $\pm$ 0.005\\
0.145   &   3.411 $\pm$ 0.01   &   0.57    &   0.754 $\pm$ 0.005\\
0.16    &   3.238 $\pm$ 0.01   &   0.64    &   0.664 $\pm$ 0.005\\
0.175   &   3.081 $\pm$ 0.01   &   0.71    &   0.581 $\pm$ 0.005\\
0.19    &   2.956 $\pm$ 0.01   &   0.78    &   0.513 $\pm$ 0.005\\
0.205   &   2.908 $\pm$ 0.01   &   0.85    &   0.456 $\pm$ 0.005\\
0.22    &   2.85 $\pm$ 0.01    &   0.92    &   0.408 $\pm$ 0.005\\
0.235   &   2.554 $\pm$ 0.01   &   0.99    &   0.366 $\pm$ 0.005\\
0.25    &   2.327 $\pm$ 0.01   &   1.06    &   0.331 $\pm$ 0.005\\
0.265   &   2.159 $\pm$ 0.01   &   1.13    &   0.299 $\pm$ 0.005\\
0.28    &   2.016 $\pm$ 0.004   &   1.20    &   0.272 $\pm$ 0.005\\
0.295   &   1.886 $\pm$ 0.004   &   1.27    &   0.247 $\pm$ 0.005\\
0.31    &   1.766 $\pm$ 0.003   &   1.35    &   0.225 $\pm$ 0.005\\
0.325   &   1.655 $\pm$ 0.003   &   1.42    &   0.205 $\pm$ 0.005\\
0.34    &   1.551 $\pm$ 0.003   &   1.49    &   0.187 $\pm$ 0.005\\
0.355   &   1.454 $\pm$ 0.002   &   1.56    &   0.171 $\pm$ 0.005\\
0.37    &   1.364 $\pm$ 0.002   &   1.63    &   0.156 $\pm$ 0.005\\
0.385   &   1.281 $\pm$ 0.002   &   1.70    &   0.142 $\pm$ 0.005\\
0.4     &   1.204 $\pm$ 0.003   &   1.77    &   0.13 $\pm$ 0.005\\
0.415   &   1.135 $\pm$ 0.003   &   1.84    &   0.118 $\pm$ 0.005\\
0.43    &   1.072 $\pm$ 0.003   &   1.91    &   0.107 $\pm$ 0.005\\
0.45    &   1. $\pm$ 0.004  &   1.98    &   0.097 $\pm$ 0.005\\
0.46    &   0.966 $\pm$ 0.004   &   2.05    &   0.088 $\pm$ 0.005\\
0.475   &   0.924 $\pm$ 0.005   &   2.12    &   0.079 $\pm$ 0.005\\
0.49    &   0.888 $\pm$ 0.005   &   2.2     &   0.071 $\pm$ 0.005\\
\enddata
\end{deluxetable}

\begin{figure*}
   \centering
    \includegraphics[width=\linewidth]{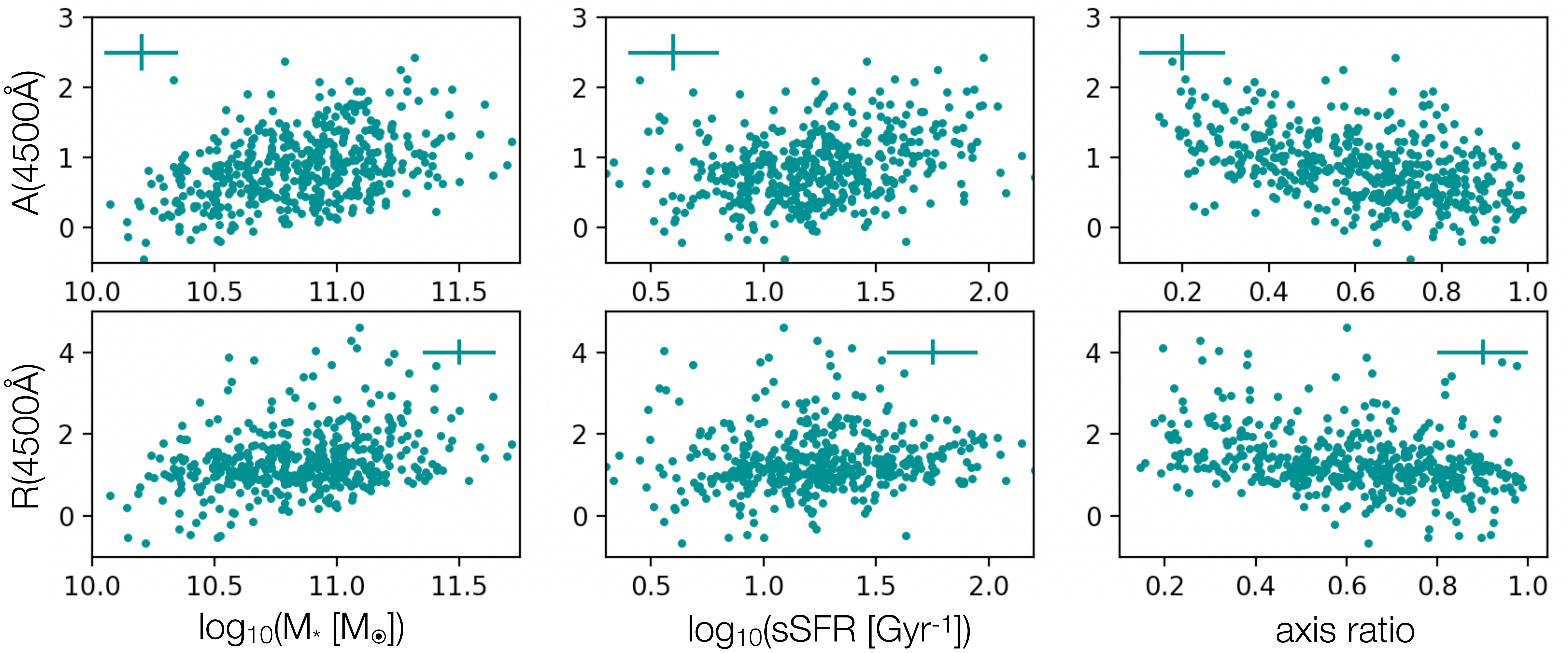}
    \caption{Attenuation A(4500\AA) (top) and the slope R(4500\AA) (bottom) as a function of global galaxy parameters. A(4500\AA) increases with stellar mass and sSFR and for edge-on galaxies. The slope R(4500\AA) on the other hand shows weak trend with global galaxy properties, on average being shallower for more massive and edge-on galaxies.}
    \label{fig:arbmsq}
\end{figure*}

\subsection{Estimating Random Uncertainties of the Attenuation Parameters}
\label{error}

The uncertainties on A(4500\AA), R(4500\AA) and B are not dominated by the photometric uncertainties used to constrain the parameters, but rather by the uncertainties in the attenuation-free stellar spectrum. The spectral fitting approach described in Section 2.2 does not allow for a straightforward estimate of this uncertainty and we resort to a `trick' that allows us to determine an upper limit on the typical random uncertainties in A(4500\AA), R(4500\AA) and B.
From the 485 galaxies with good fits we select those 113 galaxies with measured H$\delta$, D$_n$4000 index and 1 $<$ R $<$ 1.4 (the typical range) and for each galaxy we select a partner galaxy that is matched in redshift and specific SFR, as well as the spectroscopic indices H$\delta$, D$_n$4000 \citep{wu18}. The underlying assumption is that the partners have the exact same attenuation-free spectrum, and the difference between the two inferred attenuation-free spectra reflect the random uncertainty. This is obviously an exaggeration, and the matched galaxies are expected to have different stellar populations in reality. Nevertheless, this allows for a second attenuation curve fit for each galaxy (the photometry is not interchanged) and a second set of values for A(4500\AA), R(4500\AA) and B. The distribution of the differences between these sets of measurements reflect the (exaggerated) random uncertainty. After two iterations of 3 $\sigma$ rejection (accounting for the most egregious cases of mismatched attenuation-free spectra) we find standard deviations of \big\{$\sigma$(A(4500\AA)), $\sigma$(R(4500\AA)), $\sigma$(B)\} = \big\{0.26, 0.31, 0.07\}. We adopt these values as the typical random uncertainty. The dynamic range in the parameters is much larger than these uncertainty upper limits, indicative of the high level of precision of our measurements.

\subsection{Global Attenuation Properties}
\begin{figure}
    \centering
    \includegraphics[width=\linewidth]{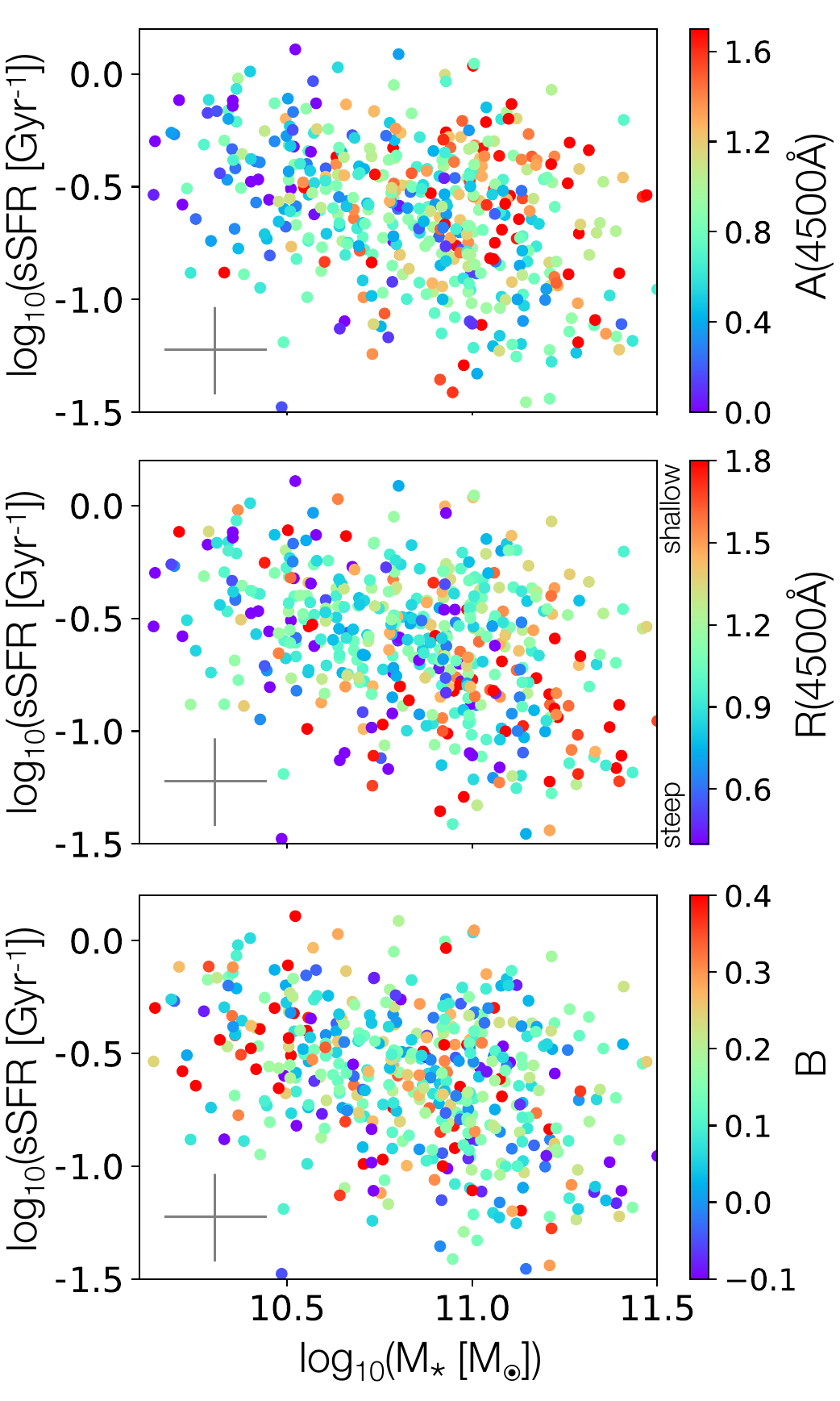}
    \caption{Specific star-formation rate (sSFR) vs. stellar mass M$_*$ for star-forming galaxies color-coded by A(4500\AA) (top panel), slope R(4500\AA) (see Eq. \ref{slope}; middle panel) and the UV bump strength B (see Eq. \ref{bump} bottom panel).} 
    \label{fig:SM}
\end{figure}

\begin{figure*}
    \centering
    \includegraphics[width=\linewidth]{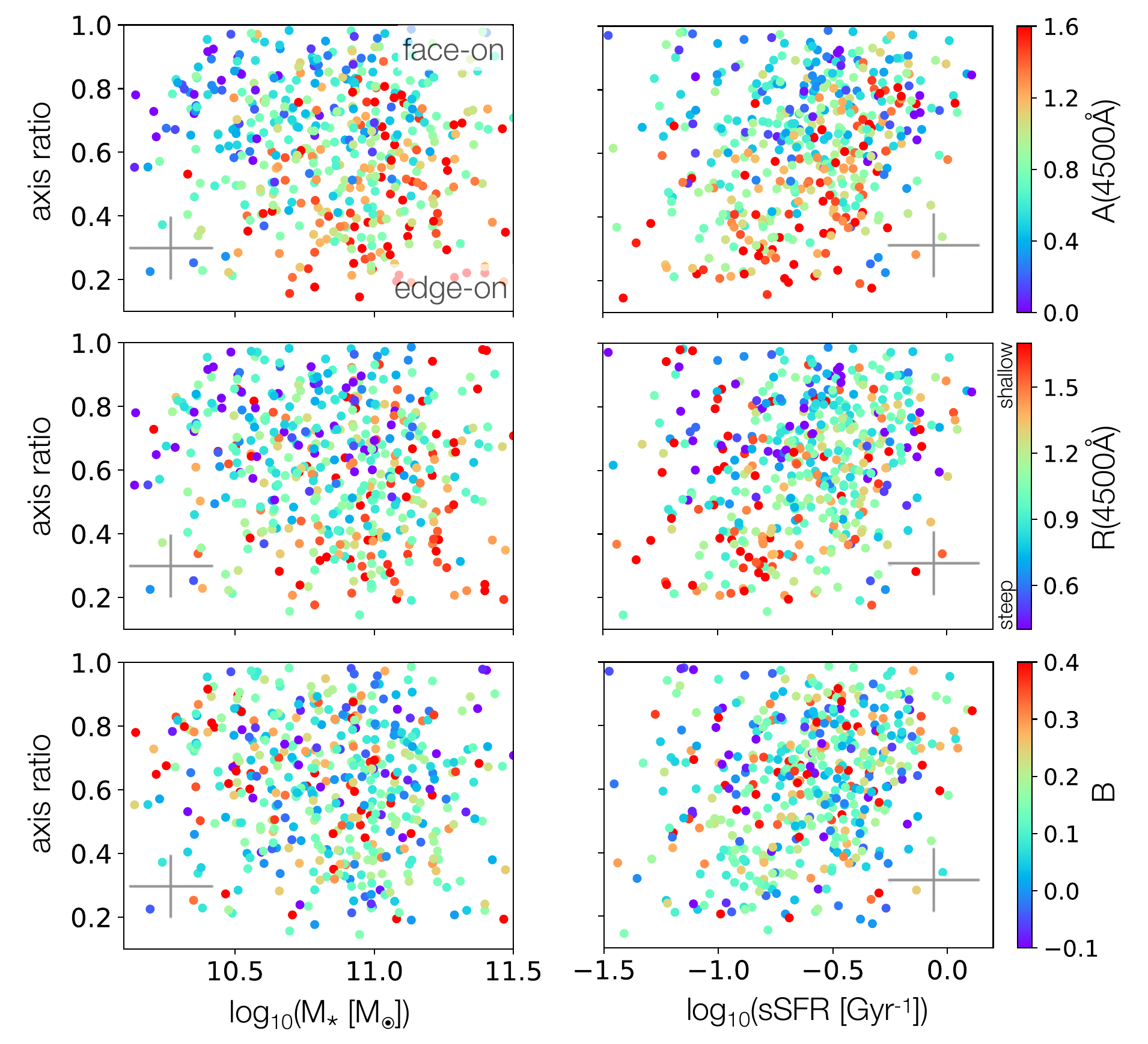}
    \caption{Axis ratio vs. stellar mass M$_*$ (left panels) and specific star-formation rate (sSFR, right panels), color-coded by the attenuation at 4500\AA\ (top panels), slope of the attenuation curve R(4500\AA) (middle panels), and strength of the UV bump feature (bottom panels). Face-on galaxies are less attenuated compared to edge-on galaxies, and also tend to have steeper slopes than edge-on galaxies.}
    \label{fig:AR}
\end{figure*}

In Figures \ref{fig:arbmsq}, \ref{fig:SM} and \ref{fig:AR} we present the variation of the dust attenuation properties of galaxies -- namely attenuation A(4500\AA), slope of the attenuation curve R(4500\AA) and UV bump strength $B$ with several global galaxy properties: specific star-formation rate (sSFR = SFR/M$_*$), stellar mass M$_*$ and projected axis ratio (as an inclination proxy).

We observe an increase in A(4500\AA) with stellar mass M$_*$, and sSFR (see Figures \ref{fig:arbmsq} and \ref{fig:SM}).
This trend can be explained by the larger dust content in more star-forming and more massive galaxies 
 -- a trend also seen in present day universe \citep[e.g.][]{cortese12, ciesla14, orellana17}, which can be attributed to the combination of large gas fractions and high metallicity \citep[e.g.][]{tremonti04, draine07, remy14}. However, the strongest correlation is between A(4500\AA) and axis ratio (see top right panel in Figure \ref{fig:arbmsq} and top panels in Figure \ref{fig:AR}), which immediately tells us that geometry to a large extent shapes observed attenuation curves, both for present day galaxies \citep[e.g.][]{salim18} and for galaxies at large lookback time \citep[see also][]{patel12, wang18}

\begin{figure*}
    \centering
    \includegraphics[width = \linewidth]{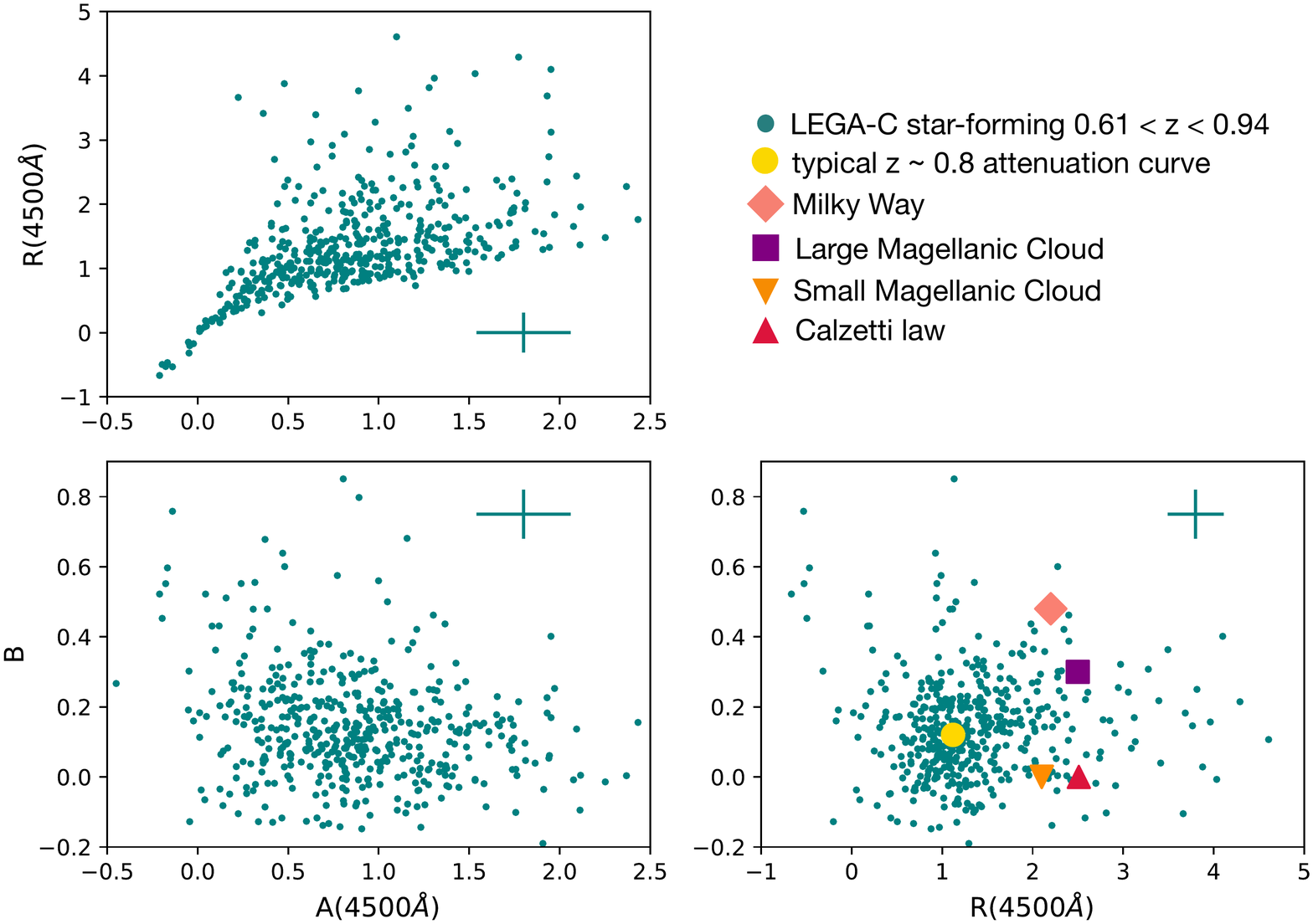}
    \caption{Attenuation curve features -- attenuation A(4500\AA), slope R(4500\AA) and UV bump strength B, plotted against each other. The explicit dependence of R(4500\AA) on A(4500\AA) makes for highly structured patterns in the distribution of these two parameters. Additionally, for comparison we also show slope and bump strength values for Milky Way, Large and Small Magellanic Clouds, translated to our parametrization. Even though left panels imply shallower slopes and weaker bump strengths with increased attenuation, there is no clear correlation between bump strength and the slope.}
    \label{fig:arb}
\end{figure*}

\begin{figure}
    \centering
    \includegraphics[width=\linewidth]{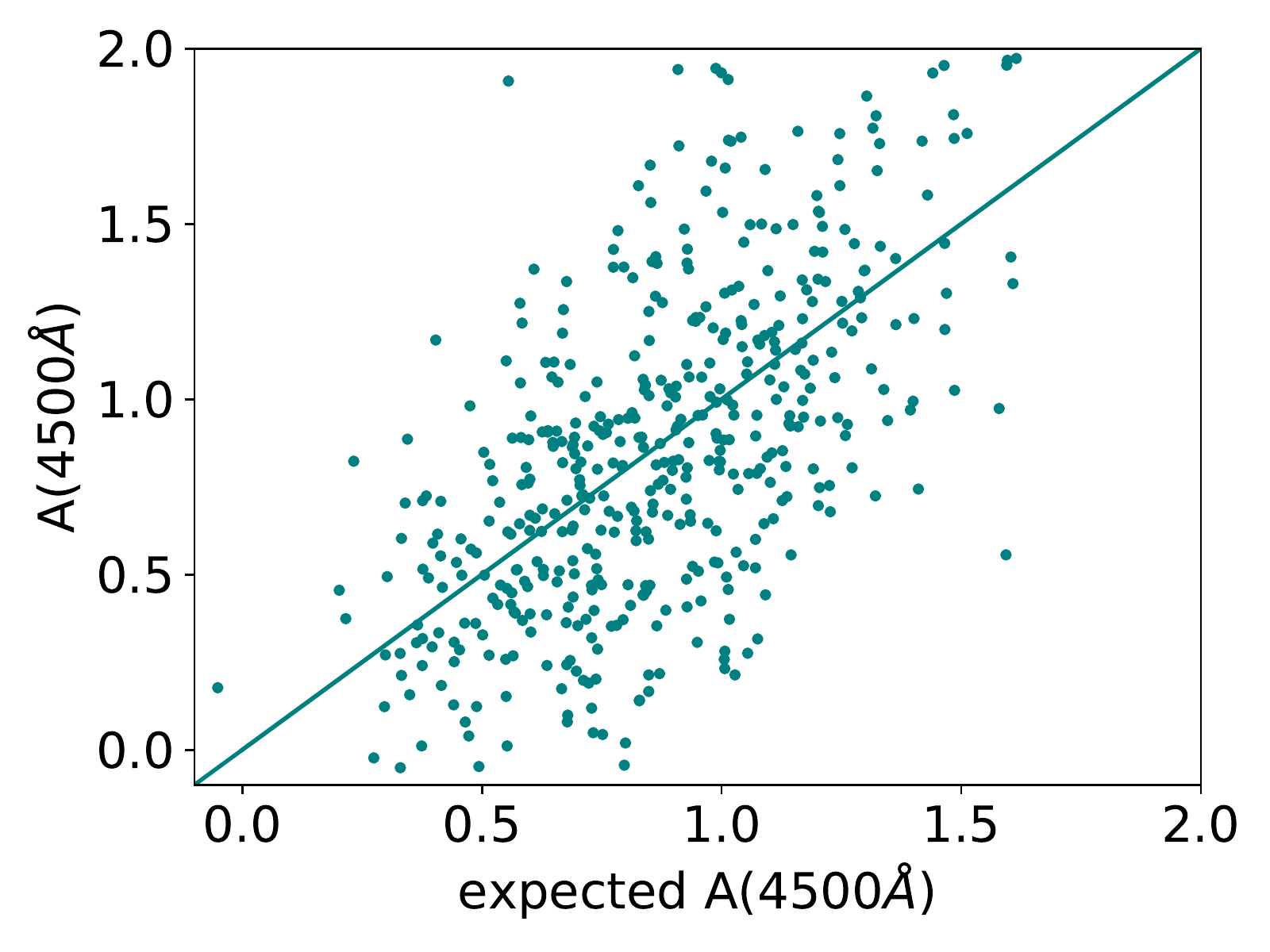}
    \caption{Expected attenuation A(4500\AA) based on the prescription given in Section \ref{sec33} as a function of the observed attenuation A(4500\AA). The diagonal line presents just a one-to-one relation. A large residual scatter still remains.}
    \label{fig:stat}
\end{figure}

\begin{figure*}
    \centering
    \includegraphics[width=\linewidth]{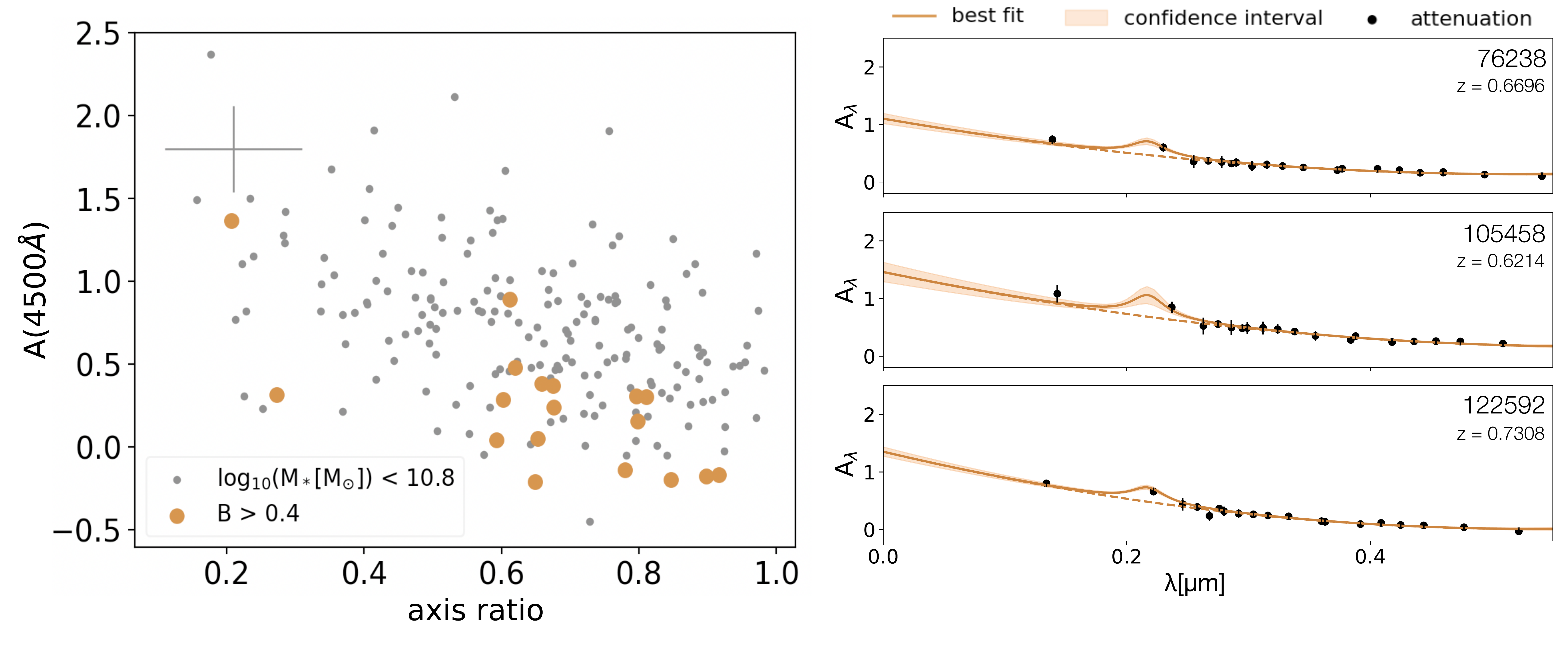}
    \caption{\textbf{Left}: Attenuation A(4500\AA) as a function of axis ratio for a sub-sample of low stellar mass galaxies (M$_*$ $<$ 10.8 M$_{\odot}$) (grey circles) with indicated galaxies with strong UV bump (orange circles). \textbf{Right:} Example attenuation curves for low stellar mass face-on galaxies with strong UV bump. A large residual scatter still remains.}
    \label{fig:faceon}
\end{figure*}

The slope R(4500\AA) does not exhibit a clear correlation with stellar mass M$_*$ or sSFR (Figure \ref{fig:SM}), but it is interesting to note that 
all of the attenuation curves of our selected sample of star forming galaxies are steeper in comparison to the classical curves in the optical wavelength range.
Our typical attenuation curve slope value (median: R(4500\AA) $\sim$ 1.2 $\pm$ 0.31) is much steeper than the MW (R(4500\AA) $\sim$ 2.2), Calzetti (R(4500\AA) $\sim$ 2.4), LMC (R(4500\AA) $\sim$ 2.5)  and SMC (R(4500\AA) $\sim$ 2.1) extinction curve (see Figure \ref{fig:arb}). The UV bump is typically much weaker than the MW UV bump (0.12 $\pm$ 0.07, compared to 0.48 for MW bump), and only a few galaxies (7\%) have B $>$ 0.4, comparable to the MW. 

We do not observe a clear trend in the sSFR - M$_*$ plane with the UV bump strength (see Figure \ref{fig:SM}), but we see variations from B = 0 to B = 0.4, indicating a large intrinsic scatter (typical uncertainty on the UV bump strength $\sim$ 0.04). However, we see an indication that more (less) massive galaxies\footnote{$\log_{10}$(M$_*$/M$_{\odot}$) $>$ 10.7}  on average have shallower (steeper) attenuation curve slope and weaker (stronger) UV bump values. Our results are therefore consistent with \cite{salim18} who find shallower slope values and a general decrease in the UV bump strength at higher stellar mass M$_*$ for present-day galaxies. In addition they find a general increase in the bump strength on both sides of the main sequence, which we do not observe. This is perhaps not surprising, given the limited sample size and a large scatter in the UV bump strength values.

Secondly, we observe steeper attenuation curve slope values for face-on galaxies, as compared to edge-on galaxies (see Figures \ref{fig:arbmsq} and \ref{fig:AR}), which is in agreement with the results from \cite{wild11} for present-day galaxies. Similarly, even though \cite{salim18} find a weak trend in the slope of the attenuation curve with galaxy orientation, they see on average a steeper slope for face-on galaxies.
{We note that the comparison between our optical slope and the slope defined using modified power-law parametrization (UV slope) is not straightforward. However, as we aim to compare only overall observed trends, we refer to the result in \cite{salim20} Figure 6B which demonstrates that steep UV slope goes together with steep optical slope.}
This trend is consistent with predictions by \cite{chevallard13}, who show that scattering at low optical depth (face-on orientation) would result in the steep observed attenuation curve. 
Their work further shows that the differences in spatial distribution and dust obscuration of young and old stellar population will also play a role. If embedded, young stars are always attenuated, including face-on viewing angle, while older stars only see increased attenuation when galaxies are viewed more edge-on. Thus, edge-on galaxies will have flatter global attenuation curves than face-on galaxies \citep[see][and references therein]{chevallard13}. Additionally, we also observe a possible indication of stronger UV bumps for low-mass galaxies, in qualitative agreement with \cite{salim18}. This could be due to lower attenuation in these galaxies, as suggested by models which predict an increasing bump strength with decreasing optical depth \citep[e.g.][]{wittgordon, pierini04, panuzzo06, inoue06, seondraine} 
This will be explored further in Section \ref{33}.

It is interesting to examine mutual dependence of the attenuation curve parameters A(4500\AA), R(4500\AA) and B (see Figure \ref{fig:arb}). Upper and bottom left panels imply possible flattening of the attenuation curve and, on average, weakening of the bump strength, as the attenuation increases. However, a possible correlation between the slope and the bump strength remains very challenging to recognize.
The highly structured patterns in the A(4500\AA) vs.~R(4500\AA) distribution are driven by the parameterizations of $A(\lambda)$ (a 2nd order polymial) and R(4500\AA).  Per Eq.~1, we have that $A(4500\AA) / A(3000\AA) = (4500/3000)^2 a + (4500/3000) b +c$, such that the slope can be written as $1/R(4500\AA) = 2.25a + 1.5b + c - 1$. The result is that variations in A(4500\AA) (or A(3000\AA)) will move $1/R(4500\AA)$ along this parabola. The only physically meaningful feature in the diagram is the lack of extremely steep ($R\lesssim 0.8$) attenuation curves for galaxies with significant attenuation at A(4500\AA). Other features and correlations are imposed by the parameterization and artificial in nature.


\subsection{Intrinsic, Unexplained Scatter in Attenuation}
\label{sec33}

So far we explored the correlation between dust attenuation curve properties and global physical properties of galaxies. Here we explore how these correlations explain the observed variety in attenuation properties: do they account for most of the variation, or are there additional variables?
To quantify this we perform a multiple linear regression to predict the attenuation at 4500$\AA$ using the following prescription:

\begin{equation}
    \nonumber A(4500\AA) = c_0 + c_1\cdot\frac{b}{a} + c_2\cdot\log_{10}(SFR) + c_3\cdot\log_{10}(M_*)
\end{equation}

\noindent and we find best fitting parameters: \big\{$c_0, c_1, c_2, c_3$\} = \big\{-2.963 $\pm$ 0.71, -1.101 $\pm$ 0.09, 0.315 $\pm$ 0.05, 0.379 $\pm$ 0.07 \}.
In Figure \ref{fig:stat} we show the expected attenuation as a function of the observed attenuation A(4500$\AA$). 
The scatter of the observed A(4500\AA) is $\sigma$ = 0.5. The formal uncertainty on A(4500\AA) is $\sigma$ = 0.26 which yields an intrinsic scatter of $\sigma$ = 0.43.
After applying the prescription above, the residual scatter in A(4500\AA) is $\sigma$ = 0.4, only somewhat reduced with respect to the observed scatter. Subtracting the contribution of the uncertainty on A(4500\AA) leaves us with the intrinsic scatter of $\sigma$ = 0.3. This means that most of the variation in A(4500\AA) is not accounted for by the known correlations with global galaxy properties. In terms of variance, 51\% of the variation is produced by these correlations, while 49\% is not accounted for.

It is worth noting that no significant contribution from other global physical properties was found -- including redshift, galaxy age, size and other structural parameters.
The reason that the scatter remains relatively large might be due to underestimated measurement uncertainties as they are dependent on the spectral models and the possible existence of unknown systematics in the photometry. However, it is also clear that we do not have a complete understanding of which physical mechanisms drive the variation in attenuation. We speculate that short-term processes ($\sim$ 10$^{8}$ yr) can randomize dust geometry and the distribution of young stars, which could play an important role. 
Up until now, at least in high redshift attenuation studies, it has not been clear whether the observed scatter in attenuation is due to measurement uncertainties, correlations with other global galaxy parameters or other. This is the first study yet to measure attenuation A(4500\AA) with sufficient precision to provide such information.

\subsection{Strong UV Bumps in Face-on Galaxies}
\label{33}

At lower stellar mass values ($\log_{10}$(M$_*$/M$_{\odot}$) $<$ 10.8) we see an increased number of low-attenuation, face-on galaxies with very strong UV bump values (see Fig \ref{fig:faceon}); this is consistent with the idea that the UV bump is associated with small dust grains which mostly reside immediately around star-forming regions\footnote{One of the suggested carriers of the UV bump itself is due to the presence of amorphous hydrocarbon materials \citep[e.g.][]{jones13}}. 
Recent model-based studies have as well shown the presence of strong UV bumps at low optical depth \citep[e.g.][]{pierini04, inoue06, seondraine}.
Why some galaxies show strong bumps and others do not -- even if similar in all other properties -- remains unclear.

\cite{salim18} found that a weaker UV bump feature is in general associated with shallower attenuation curve slope values. Moreover, even high redshift studies come to similar conclusions: specifically \cite{kriek13} find that their attenuated sub-sample of A$_{V}$ $>$ 0.5 galaxies shows stronger UV bump feature when the slope of the attenuation curve is steeper. 
We can not draw this conclusion directly from Figure \ref{fig:SM} as a strong UV bump feature is in general not associated with particularly steep or shallow slope values.
In general we do not see a clear trend between the UV bump strength and the slope. Given the uncertainties on the UV bump and slope measurement which are approximately the size of the scatter, it is difficult to draw conclusions regarding the existence of an underlying trend.

Even though the above mentioned studies have found evidence for a correlation between the slope of the attenuation curve and the bump strength, 
we do not find such a trend in this work.

The choice of the attenuation curve parametrization and of the UV bump profile dictates the measurement of the UV bump strength and of the slope of the attenuation curve, due to the fact that they are not independent from each other.
The variety of attenuation curve parametrization choices in the literature makes the comparison between studies rather difficult. Additionally, a wide variety in procedures involved in obtaining the attenuation information makes for a challenging comparison as well. Taking this into account, diversity in prescriptions among studies and techniques applied to derive the attenuation may explain inconsistencies in terms of the trends between the slope of the attenuation curve, strength of the bump feature and general attenuation with other galaxy properties \citep[e.g.][]{buat12, battisti16, battisti17, kriek13, salim18}.

Still, it seems unlikely that we can attribute the difference to the parametrization choice, given the agreement between some local based \citep[e.g.][]{salim18} and high-redshift based \citep[e.g.][]{kriek13} studies in terms of trends between the slope of the attenuation curve and the strength of the UV bump -- regardless of them applying mutually distinct modified Calzetti dust law parametrization. Even the high redshift low-resolution study based on individual galaxies by \cite{tress18} following higher order polynomial based parametrization \citep{conroy10} find an indication of a similar dependence between the slope of the attenuation curve and the UV bump strength. On the other hand, a high redshift study by \cite{buat12}, who apply the same modified Calzetti dust law parametrization as \cite{noll09} do not find any correlation between the slope of the attenuation curve and the UV bump strength. Similarly, \cite{burgarella05} finds no particular underlying trend between the slope of the attenuation curve and UV bump strength for their local galaxy samples.

Possible differences between our and other studies may emerge from the methods applied in deriving the attenuation. Determination of the attenuation at high redshift in past studies have originated from the SED fits based on the observed multi-band photometry \citep[e.g.][]{buat12} and often rely on stacks of the derived SEDs in order to achieve the equivalent-to low-resolution observed spectra \citep[e.g.][]{kriek13}. 
We note that these previous studies do not use high signal-to-noise spectroscopy to infer the underlying stellar continuum, which is a clear advantage of this study.
The main caveat of our study is that we do not take into account any systematic uncertainties in the stellar population model used to model our spectra. Future work will include a systematic exploration of several stellar population models and the effect of that choice on the inferred attenuation.

\section{Conclusion}
We explore the attenuation curve diversity of individual star-forming galaxies at 0.61 $<$ z $<$ 0.94. Our results are based, for the first time, on deep and high-resolution optical spectra. Using this unique information we provide a new attenuation curve prescription for z $\sim$ 0.8 star-forming galaxies. Based on the broad agreement of our results with the results from \cite{salim18} on present-day galaxies we suggest that our new prescription (Equations \ref{prescription} and \ref{modified}) is generally more accurate than commonly adopted prescriptions \citep[e.g.][]{calzetti00, cardelli89} to model the integrated SEDs of galaxies. We show that these galaxies exhibit a wide variety of properties in terms of the slope of the attenuation curve R(4500\AA) and strength of the UV bump feature. 
Since the significant amount of scatter in measured attenuation parameters (e.g. A(4500\AA)) is not accounted for, we argue that we can not provide the attenuation curve prescription based on global galaxy parameters (M$_*$, SFR, and b/a).

We explore the dependence of the attenuation curve features on the galaxy orientation, stellar mass M$_*$ and sSFR. In general, we observe steeper attenuation curve slopes as compared to the MW, SMC and LMC, and even the Calzetti attenuation curve. We observe that the UV bump strength is weaker as compared to the MW bump strength value ($\sim$ 0.5).
We see an increase in overall attenuation A(4500\AA) with sSFR and stellar mass M$_*$, however, any underlying trend with the slope of the attenuation curve and the UV bump strength is very challenging to recognize.
The strongest correlation we find is between the galaxy orientation and the overall attenuation and slope. We find that face-on galaxies are less attenuated and have a steeper slope -- suggesting the geometry has a significant influence on the observed attenuation. Secondary correlations with stellar mass and star-formation rate can be attributed to the higher metallicities and dust masses of higher-mass, gas-rich galaxies. Additionally, we also observe an indication of a stronger bump feature in face-on low stellar mass M$_*$ galaxies.
Despite the systematic changes of dust attenuation with global galaxy properties and viewing angle,
most of variations in attenuation are not accounted for.
This implies that more subtle geometric effects and/or random factors such as patchiness of the young stellar populations and dust patterns dominate the attenuation.

\section{Acknowledgements}
We thank the anonymous referee for providing valuable feedback. 
CP is supported by the Canadian Space Agency under a contract with NRC Herzberg Astronomy and Astrophysics.
This project has received funding from the European Research Council (ERC) under the European Union's Horizon 2020 research and innovation programme (grant agreement 683184). Based on observations made with ESO Telescopes at the La Silla or Paranal Observatories under programme ID 194.A-2005.

\bibliographystyle{aasjournal}
\bibliography{bibtex.bib}

\end{document}